\documentclass[epj]{webofc}
\usepackage[varg]{txfonts}   
\doi{}
\usepackage{subfig}

\usepackage{hyperref}
\hypersetup{
  pdftitle={},%
  pdfauthor={},%
  pdfsubject={},%
  pdfkeywords={},%
  pdfstartview={},%
  bookmarksopen=true, breaklinks=true, debug=true, %
  colorlinks=true, linkcolor=blue, citecolor=blue, urlcolor=blue
}

\newcommand{\glabcms}{\gamma^{\rm lab}_{\rm c.m.s.}}
\newcommand{\blabcms}{\beta^{\rm lab}_{\rm c.m.s.}}
\newcommand{\dylabcms}{\Delta y^{\rm lab}_{\rm c.m.s.}}

\newcommand{\ie}{{\it i.e.}}
\newcommand{\eg}{{\it e.g.}}
\newcommand{\etal}{{\it et al.}}

\wocname{EPJ Web of Conferences}
\woctitle{INPC 2013}
\begin{document}
\title{AFTER@LHC: a precision machine to study the interface between particle and nuclear physics}
%
%

\author{
J.P. Lansberg\inst{1} \thanks{\email{Jean-Philippe.Lansberg@in2p3.fr}}
\and
R.~Arnaldi\inst{2}
\and
S.J.~Brodsky\inst{3}
\and
V.~Chambert\inst{1}
\and
J.P. Didelez\inst{1}
\and
B. Genolini\inst{1}
\and
E.G.~Ferreiro\inst{4}
\and
F.~Fleuret\inst{5}
\and
C.~Hadjidakis\inst{1}
\and
C.~Lorc\'e\inst{1}
\and
A.~Rakotozafindrabe\inst{6}
\and
P.~Rosier\inst{1}
\and
I.~Schienbein\inst{7}
\and
E.~Scomparin\inst{2}
\and
U.I.~Uggerh\o j\inst{8}
}

\institute{
IPNO, Universit\'e Paris-Sud, CNRS/IN2P3, F-91406, Orsay, France
\and
Dip. di Fisica and INFN Sez. Torino, Via P. Giuria 1, I-10125, Torino, Italy      
\and
SLAC National Accelerator Lab., Theoretical Physics, Stanford University, Menlo Park, CA 94025, USA  
\and
Departamento de F{\'\i}sica de Part{\'\i}culas, Universidade de Santiago de C., 15782 Santiago de C., Spain
\and
Laboratoire Leprince Ringuet, \'Ecole Polytechnique, CNRS/IN2P3,  91128 Palaiseau, France
\and
IRFU/SPhN, CEA Saclay, 91191 Gif-sur-Yvette Cedex, France%
\and
LPSC, Universit\'e Joseph Fourier, CNRS/IN2P3/INPG, F-38026 Grenoble, France
\and
Department of Physics and Astronomy, University of Aarhus, Denmark 
}

\abstract{%
We outline the opportunities to study with high precision the interface between
nuclear and particle physics,
which are offered by a next generation and multi-purpose fixed-target experiment exploiting 
the proton and ion LHC beams extracted by a bent crystal. 
}
\maketitle
\vspace*{-.2cm}
\section{Introduction}
\label{intro}

With the advent of RHIC and the LHC, collisions of heavy ions can nowadays be studied 
at ultra-relativistic energies where phenomena such as the creation of quark-gluon plasma,
 the saturation of gluon densities at low $x$, or even heavy-flavour regeneration can be 
studied. Clearly, this has opened new horizons to study heavy-ion physics in the high-energy 
limit, one of the interface between particle and nuclear physics.

There are however other very interesting kinematical regions at the interface 
between particle and nuclear physics which have been seldom 
explored and from which one can learn much on the dynamics of the strong interaction. 
The domain of large momentum fractions at small distances, which can be studied by means of hard 
reactions in the backward region in the fixed-target collision mode, is particularly interesting.

In this context, it is useful to recall the critical advantages of a fixed-target experiment 
compared to a collider one, \ie~
\begin{itemize} \itemsep-4pt
\item[-] extremely high luminosities thanks to the high density of the target; 
\item[-] absence of geometrical constraints to access the far backward region in the c.m.s.;
\item[-] unlimited versatility of the target species; and
\item[-] same energy for proton-proton, proton-deuteron and proton-nucleus collisions.
\end{itemize}

Indeed, high luminosities allow one to look for rare configurations, which are however important to 
understand the dynamics of the strong interaction. The study of reactions in the target-rapidity 
region -- the far backward region-- allows one to probe large momentum fractions $x$ in the target, exactly at the interface 
between the dynamics of the nucleons and the nucleus. The versatility of the target allows one to systematically study the 
mass or size dependence of the nuclear effects. Finally, the possibility to study
proton-proton and proton-deuteron collisions at the same energy and with the same experimental conditions 
allows one to extract out in the most straighforward fashion the effects only due to the collective dynamics of the 
nucleons in the nucleus.

\section{Luminosities for a bent-crystal-extraction mode on the LHC beams}

To our knowledge, the most convenient and efficient way to obtain a beam for a fixed target 
experiment from the LHC is to use bent-crystal beam extraction. The idea is therefore to
position a bent crystal in the halo of the beam such that a few protons (or lead) per bunch
per pass would be channelled in the lattice of the crystal and would be deviated by a couple of
mrad w.r.t. to the axis of the beam. Such a method also has the virtue of better collimating  the beam,
allowing one to increase the luminosity of the collider experiments.

The extraction of half of the LHC proton beam loss, \ie~$5\times 10^8$ p$^+/s$, would be high enough 
to perform $pp$, $pd$ and $pA$  collisions with luminosities way above that of RHIC 
in the same energy range. Indeed, 7 TeV protons colliding 
on fixed targets release a center-of-mass energy close to 115 GeV ($\sqrt{2E_p m_N}$).
As regards heavy-ion beams, in particular lead ions, tests to extract the lead beam 
of SPS have been successfully carried out  by the CERN UA9 
collaboration~\cite{Scandale:2011zz}. In this case, the energy in the c.m.s. reaches 72 GeV per 
nucleon-nucleon collisions with the 2.76 TeV lead beam.

Experiments have demonstrated that the crystal degradation is about $6\%$ per 
$10^{20}$ particles/cm$^2$, see \eg~\cite{Baur00}. This corresponds to approximately 
a year of operation, for realistic impact parameters and beam sizes at the 
location of the crystal. After a year, the crystal has to be moved by less than a 
millimeter to let the beam halo hit the crystal  on an intact spot. Such a procedure can
be repeated almost at will.

\begin{table}[!hbt]
\scriptsize 
\centering
\subfloat[]{\begin{tabular}{cccccc}
\hline
Target       & $\rho$        &$A$ & $\mathcal L$                     & $\int \mathcal L$\\
(1 cm thick) & (g cm$^{-3}$) &    & ($\mu$b$^{-1}$ s$^{-1}$) & (pb$^{-1}$ yr$^{-1}$)\\
\hline
solid H  & 0.088 & 1   & 26 & 260 \\
liquid H & 0.068 & 1   & 20 & 200 \\
liquid D & 0.16  & 2   & 24 & 240 \\
Be       & 1.85  & 9   & 62 & 620 \\
Cu       & 8.96  & 64  & 42 & 420 \\
W        & 19.1  & 185 & 31 & 310 \\
Pb       & 11.35 & 207 & 16 & 160 \\
\hline
\end{tabular} \label{tab:lumi-pA}~}
\subfloat[]{\begin{tabular}{ccccc}
\hline
Target&$\rho$&$A$&$\mathcal L$ &$\int \mathcal L$ \\
(1 cm thick) &(g cm$^{-3}$)&&(mb$^{-1}$ s$^{-1}$)&(nb$^{-1}$ yr$^{-1}$)\\
\hline 
solid H  & 0.088 & 1   & 11 & 11 \\
liquid H & 0.068 & 1   & 8  & 8  \\
liquid D & 0.16  & 2   & 10 & 10 \\
Be       & 1.85  & 9   & 25 & 25 \\
Cu       & 8.96  & 64  & 17 & 17 \\
W        & 19.1  & 185 & 13 & 13 \\
Pb       & 11.35 & 207 & 7  & 7  \\
\hline
\end{tabular}\label{tab:lumi-PbA}}
\caption{Instantaneous and yearly luminosities obtained for various 1cm thick targets 
with an extracted beam of (a)
$5 \times 10^8$ p$^+$/s with a momentum of 7 TeV and (b) $2 \times 10^5$ Pb/s with a momentum per nucleon of 2.76 TeV.}
\label{tab:lumi}\vspace*{-.5cm}
\end{table}

Recently, the LHC Committee has expressed that "...it may well be 
feasible to bend even the low emittance LHC beam" and that "possible future applications 
of the bent-crystal scheme abound: including beam-halo cleaning and \emph{slow extraction}" 
[our emphasis]. Thus, the LHCC has recommended further studies to be performed at the 
LHC~\cite{LHCC107}, in particular those followed the CERN LUA9 collaboration for  a "smart collimator" solution, 
originally proposed in~\cite{Biryukov:2003ys}.  
Another possibility is the proposal~\cite{Uggerhoj:2005xz} to "replace" the kicker-modules in LHC section IR6, where the beam is dumped, by a bent crystal. This would give a sufficient kick to some particles in the beam halo to overcome the septum blade and then to be extracted.

The instantaneous and yearly (10$^7$~s) luminosities reachable
with the proton beam on various 1cm thick targets are gathered in table~\ref{tab:lumi-pA}. 
Note that 1m long targets of liquid hydrogen or deuterium
 give luminosities close to 20 femtobarn$^{-1}$.
Table~\ref{tab:lumi-PbA} gathers the corresponding values for the 
Pb run (10$^6$~s).

\section{Accessing the far backward region}

For proton runs,  the boost between c.m.s.
and the lab system is rather large, $\glabcms=\sqrt{s}/(2m_p)\simeq 60$ and the rapidity
shift is $\tanh^{-1} \blabcms\simeq 4.8$. In the lead case, one has
$\glabcms\simeq 38$ and  $\dylabcms\simeq 4.3$.
In both cases, the c.m.s. central-rapidity region, $y_{\rm cms}\simeq 0$, is thus highly boosted at an angle of about one degree w.r.t.
 the beam axis  in the laboratory frame. The entire backward c.m.s. hemisphere ($y_{\rm cms}<0$) 
is easily accessible with standard experimental techniques. The forward hemisphere is less 
easily accessible because of the reduced distance from the (extracted) beam axis which  requires the use of highly segmented 
detectors to deal with the large particle density. In a first approach, we consider that 
one can access the region $-4.8\leq y_{\rm cms}\leq 1$ without specific difficulty.
  This allows for the detection of the main part of the
particle yields as well as high precision measurements in the whole backward hemisphere, down to 
$x_F\to -1$ for a large number of systems. As an example, the study of $\Upsilon$ production at
rapidities of the order of -2.4 in the c.m.s. already provides with an access to $x_F$ above $\frac{10}{115} e^{2.4}\simeq 0.95$.
The leptonic decay products of such $\Upsilon$ would fly on average at angles still smaller than 45$^\circ$ in the laboratory frame ($\eta=0.88$).

\section{Novel physics studies at large {\boldmath $x$}  and small distances\protect\footnote{In what follows, we have made short selection of flagship studies. A more complete survey of the physics
opportunities with AFTER can be found in~\cite{Brodsky:2012vg}.}}

\vspace*{-.2cm}
\subsection{pQCD studies and PDF measurements at large {\boldmath $x$} in {\boldmath $pp$}, {\boldmath $pd$} and {\boldmath $pA$} collisions}

If AFTER is designed as a multi-purpose detector, it would definitely be --thanks to its outstanding luminosity and an easy access towards low $P_T$-- 
 a heavy-flavour, quarkonium and prompt-photon {\it observatory}~\cite{Brodsky:2012vg,Lansberg:2012kf} in 
$pp$ and $pA$ collisions.  
The production mechanisms of quarkonia~\cite{review} would be constrained thanks to the large quarkonium yields
 and precise measurements of their correlations, along with the forthcoming LHC results.

By accessing the target-rapidity region, the proton, neutron and nucleus 
gluon and heavy-quark distributions (see \eg~\cite{Diakonov:2012vb})  could then 
be accessed at mid and large momentum fractions, $x$. The physics at $x$ larger than unity  can be accessed
in the nuclear case. In the case of bottomonium production and prompt photon production, 
nuclear phenomena at distances of the order of 0.01 fm can be studied ! The study of the scale dependence
of nuclear effect in the EMC and Fermi motion region, $0.3 < x < 1$, could be of great help
in the understanding of the connexion between the EMC effect and the importance of short-range correlations.
 
In proton-deuteron collisions, unique information on the 
quarkonium production momentum distribution of the gluon in the neutron
can be also obtained with quarkonium measurements along the lines of E866 for $\Upsilon$~\cite{Zhu:2007mja}. Finally, 
linearly polarised gluons inside unpolarised protons~\cite{Boer:2012bt} can also be accessed
with the study of low-$P_T$ scalar and pseudoscalar quarkonium.

\vspace*{-.2cm}
\subsection{Deconfinement studies in {\boldmath ${\rm Pb}A$} collisions between SPS and RHIC energies}

Among the proposed observables to study deconfinement in relativistic heavy-ion collisions, 
one can certainly highlight the $J/\psi$ and $\Upsilon$ suppression, the heavy-flavour and jet quenching or the production 
of direct photons. These could easily be accessed with AFTER~\cite{Brodsky:2012vg,Rakotozafindrabe:2012ei}.
In particular, modern detection technology should  allow 
for thorough studies of the behaviour of the quarkonium excited states in the hot nuclear matter. Let us cite 
the $\chi_c$ and $\chi_b$ resonances. Such studies would benefit from the boost of the fixed-target mode,  
inspite of the challenges posed by the high-multiplicity environment of heavy-ion collisions. 
Dedicated studies of quarkonium excited states are of highest relevance to study 
the quarkonium sequential melting in the deconfined matter~\cite{Lansberg:2012kf} .

\vspace*{-.2cm}
\subsection{Ultra-peripheral
{\boldmath $pA/{\rm Pb}A$} collisions and {\boldmath $W/Z$} production in their threshold region}

The available c.m.s energy in the fixed-target mode is usually a limiting factor.  For the first time, 
ultra-peripheral $pA$/Pb$A$ collisions on fixed targets would occur at sufficient $\sqrt{s}$
to produce heavy mesons and hard di-leptons. The nucleus target/projectile $A$/Pb 
remains a coherent high-energy photon source, up to $E_\gamma \simeq 200$~GeV, allowing one 
for photon-hadron collision studies.
A great potential of studies is also offered by semi-diffractive processes due
to the absence of pile-up in such a slow extraction mode~\cite{Brodsky:2012vg,Lansberg:2012sq,Lorce:2012rn}.

In addition,  $W$ and $Z$ boson production could be studied for the first time close to threshold, thanks
to the high energy of the proton LHC beams and the outstanding luminosities of the fixed-target mode.
Investigations in the threshold region could be very important to understand QCD corrections
in this uncharted part of the phase space. $W$ and $Z$ boson could also be studied in $pA$ collisions
 possibly shedding light on nuclear effects (since $x_2>0.7$) at the weak-boson-mass scale, on the order of the attometer !

{\small {\bf Acknowledgements.}
This research [SLAC-PUB-15719] was supported in part by the French CNRS, grants PICS-06149 Torino-IPNO \& PEPS4AFTER2 and the Department of Energy, contract DE-AC02-76SF00515. 
}
\vspace*{-.2cm}

%

\begin{thebibliography}{}
%


 \bibitem{Brodsky:2012vg}
  S.~J.~Brodsky, F.~Fleuret, C.~Hadjidakis, J.~P.~Lansberg,
Phys.\ Rept.\ {\bf 522} (2013) 239.



 \bibitem{Lansberg:2012kf}
  J.~P.~Lansberg, S.~J.~Brodsky, F.~Fleuret and C.~Hadjidakis,
Few Body Syst.\  {\bf 53} (2012) 11.

 \bibitem{LUA9} W.~Scandale, {\it et al.} [LUA9], CERN-LHCC-2011-007, 2011.

\bibitem{Uggerhoj:2005xz}
  E.~Uggerh\o j, U.~I.~Uggerh\o j,
  Nucl.\ Instrum.\ Meth.\  B {\bf 234} (2005) 31.

\bibitem{Baur00} A. Baurichter \textit{et al.}, Nucl. Instr. Meth. B \textbf{164}, 27 (2000)

\bibitem{LHCC107} LHC Committee, minutes of the 107th meeting, CERN/LHCC 2011-010

\bibitem{Biryukov:2003ys}
  V.~M.~Biryukov,
  physics/0307027.

\bibitem{Arduini:1997nb}
  G.~Arduini,  {\it et al.} 
  Phys.\ Rev.\ Lett.\  {\bf 79} (1997) 4182.

\bibitem{Scandale:2011za}
  W.~Scandale,  {\it et al.} 
  Phys.\ Lett.\ B {\bf 703} (2011) 547.

\bibitem{Balling:2009zz}
  P.~Balling, {\it et al.} 
  Nucl.\ Instrum.\ Meth.\ B {\bf 267} (2009) 2952.

\bibitem{Scandale:2011zz}
  W.~Scandale, {\it et al.}, 
  Phys.\ Lett.\  {\bf B703 } (2011)  547-551.
 
\bibitem{review}
  Z.~Conesa del Valle \etal, 
  Nucl.\ Phys.\ (PS)  {\bf 214} (2011) 3;
  N.~Brambilla \etal, 
  Eur.\ Phys.\ J.\ C {\bf 71} (2011) 1534;
  J.~P.~Lansberg,
  Eur.\ Phys.\ J.\ C {\bf 61} (2009) 693 \&
  Int.\ J.\ Mod.\ Phys.\ A {\bf 21} (2006) 3857

\bibitem{Diakonov:2012vb}
  D.~Diakonov, M.~G.~Ryskin and A.~G.~Shuvaev,
  JHEP {\bf 1302} (2013) 069

\bibitem{Boer:2012bt}
  D.~Boer and C.~Pisano,
  Phys.\ Rev.\ D {\bf 86} (2012) 094007

\bibitem{Zhu:2007mja}
  L.~Y.~Zhu {\it et al.}  [E866 Coll.],
  Phys.\ Rev.\ Lett.\  {\bf 100} (2008) 062301

\bibitem{Rakotozafindrabe:2012ei}
  A.~Rakotozafindrabe, {\it et al.}, 
  Nucl.\ Phys.\ A {\bf 904-905} (2013) 957c

\bibitem{Lansberg:2012sq}
  J.~P.~Lansberg, {\it et al.}, 
PoS  {\bf ICHEP2012} (2012) 547


\bibitem{Lorce:2012rn}
  C.~Lorce, {\it et al.}, 
  AIP Conf.\ Proc.\  {\bf 1523} (2012) 149

\end{thebibliography}
%
%

\end{document}